\documentclass{llncs}

\usepackage[english]{babel}
\usepackage{alltt}
\usepackage{graphicx}
\usepackage{url}
\usepackage{subfigure}
\usepackage{listings}
\usepackage{color}
\usepackage{lscape}
\usepackage{xspace}
\usepackage{latexsym}
\usepackage{rotating} 
\usepackage{enumitem}
\usepackage{verbatim}
\usepackage{url}
\usepackage{multirow}
\usepackage{chngcntr}

\begin{document}

\title{Notes on Physical \& Logical Data Layouts}
	\author{
	Michael Hausenblas\inst{1} 
	}
	\institute{MapR Technologies EMEA, Galway, Ireland\\
	\email{mhausenblas@maprtech.com}
	}
\maketitle

\begin{abstract}
In this short note I review and discuss fundamental options for physical and
logical data layouts as well as the impact of the choices on data processing.
I should say in advance that these notes offer no new insights, that is, 
everything stated here has already been published elsewhere. In fact, it has
been published in so many different places, such as blog posts, in the 
literature, etc. that the main contribution is to bring it all together in one
place.
\end{abstract}

\pagestyle{plain}
\pagenumbering{Roman}

\section{Motivation}
\label{sec:mot}
Data processing and management systems such as databases, 
datastores~\cite{Cattell:SIGMOD11} or query engines usually have to answer 
to two kinds of entities: \emph{humans} and \emph{hardware}. 

Towards humans, they provide means to query, manipulate or 
manage\footnote{The management aspect can span a wide range of activities 
including but not limited to snapshots, mirroring, etc.} data.
Towards the hardware, they issue store and retrieve commands. They depend 
directly or indirectly on the very nature of the hardware. Almost all systems---
for example, Hadoop's distributed file system---are designed with strong though
not necessarily explicit assumptions about the underlying hardware such as hard 
disk drives (HDD)~\cite{Elerath:ACMQ1}, their spindles, heads, etc.

Conceptually, there are three levels present in data processing and management 
systems (Fig.~\ref{fig:data-layers}):
\begin{figure}[ht!]
\centering
\includegraphics[width=0.6\textwidth]{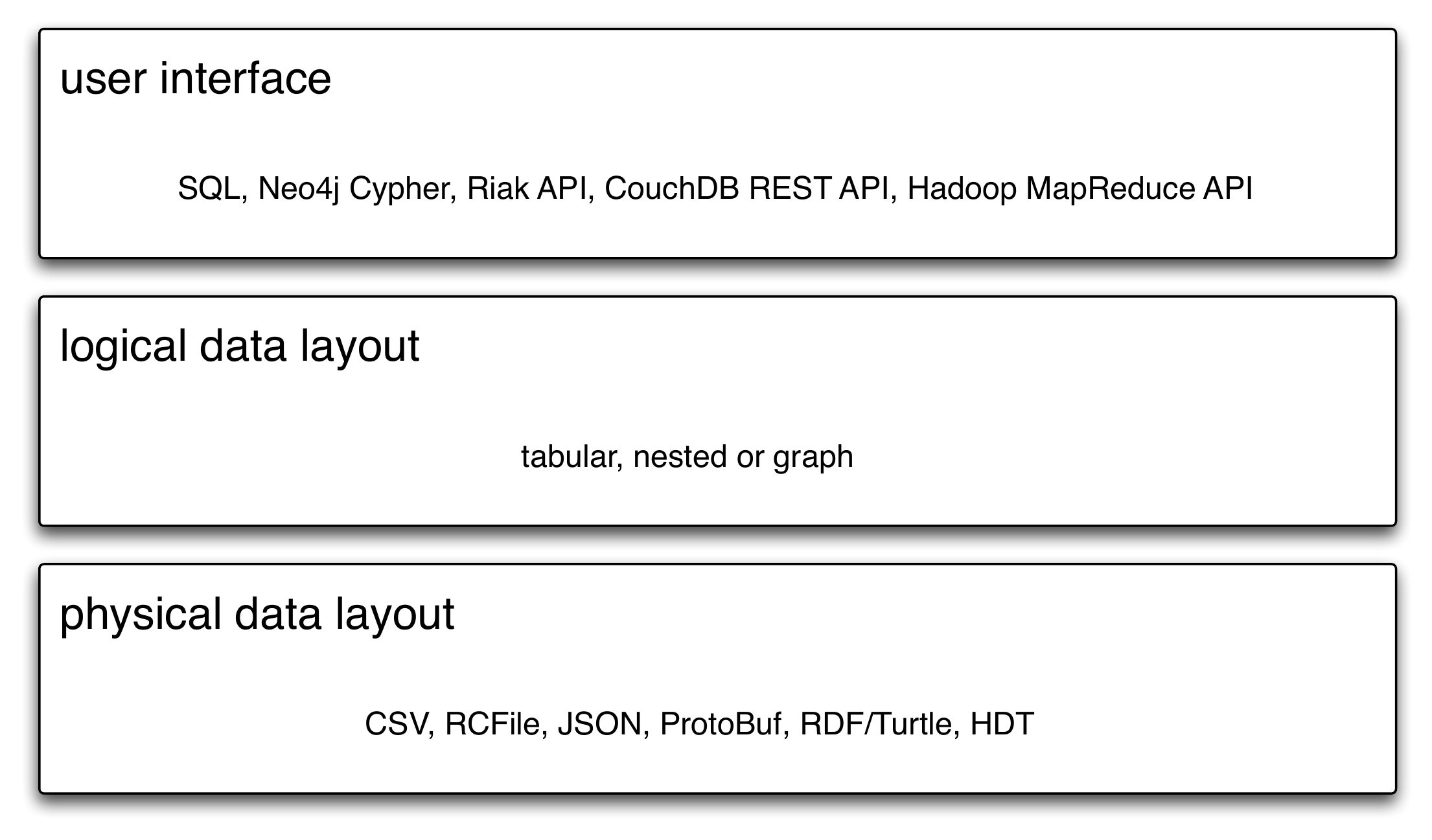}
\caption{The three levels of data representation and interaction in data 
management systems, including examples for each of the levels.}
\label{fig:data-layers}
\end{figure}
\begin{itemize}
\item The \emph{User Interface} level. Any database or datastore needs 
to provide a way to interact with the data under management. This can be 
something elaborate, standardised and mature as the Structured Query Language 
(SQL) found in relational database management systems (RDBMS), such as 
Oracle DB, PostgreSQL, or MySQL. This can be a RESTful interface, found in many 
NoSQL datastores, like, for example, CouchDB's API\footnote{See documentation at
\url{http://wiki.apache.org/couchdb/HTTP_Document_API}}. Of course, this can 
also be a programming-language-level API such as the case with
Hadoop\footnote{\url{http://hadoop.apache.org/docs/current/api/org/apache/hadoop
/mapreduce/package-summary.html}}.
\item The \emph{Logical Data Layout} level. Addresses how the user conceptually
thinks about and deals with the data. In case of a RDBMS the data units might be
tables and records, in a key-value store like Redis it may be an entry 
identified via a key and in a wide-column store the data unit might be a row
containing different columns, and last but not least in an RDF store a single
triple might be the unit one logically manipulates.
\item The \emph{Physical Data Layout} level. On this level, we're concerned with
the question how the data is laid out once serialised. The serialisation takes
place from main memory (RAM) either to send the data in question over the wire,
or, to be stored on a durable medium such as a hard disk drive or a solid-state 
drive (SSD)~\cite{Cornwell:ACMQ1}. Concrete serialisations may
be textual based, such as CSV and JSON or of binary nature, like the RCFile 
format~\cite{He:ICDE11}. 
\end{itemize}

\section{Manifestations of Data Layouts}
\label{sec:mani}
In~\cite{Hausenblas:arxiv2012} we introduced the three fundamental data shapes
\emph{tabular}, \emph{tree}---henceforth \emph{nested}, and \emph{graph}. It
turns out that it is useful to further differentiate the shapes, distinguishing
between logical and physical layouts, as hinted above.

\begin{figure}[ht!]
\centering
\includegraphics[width=0.7\textwidth]{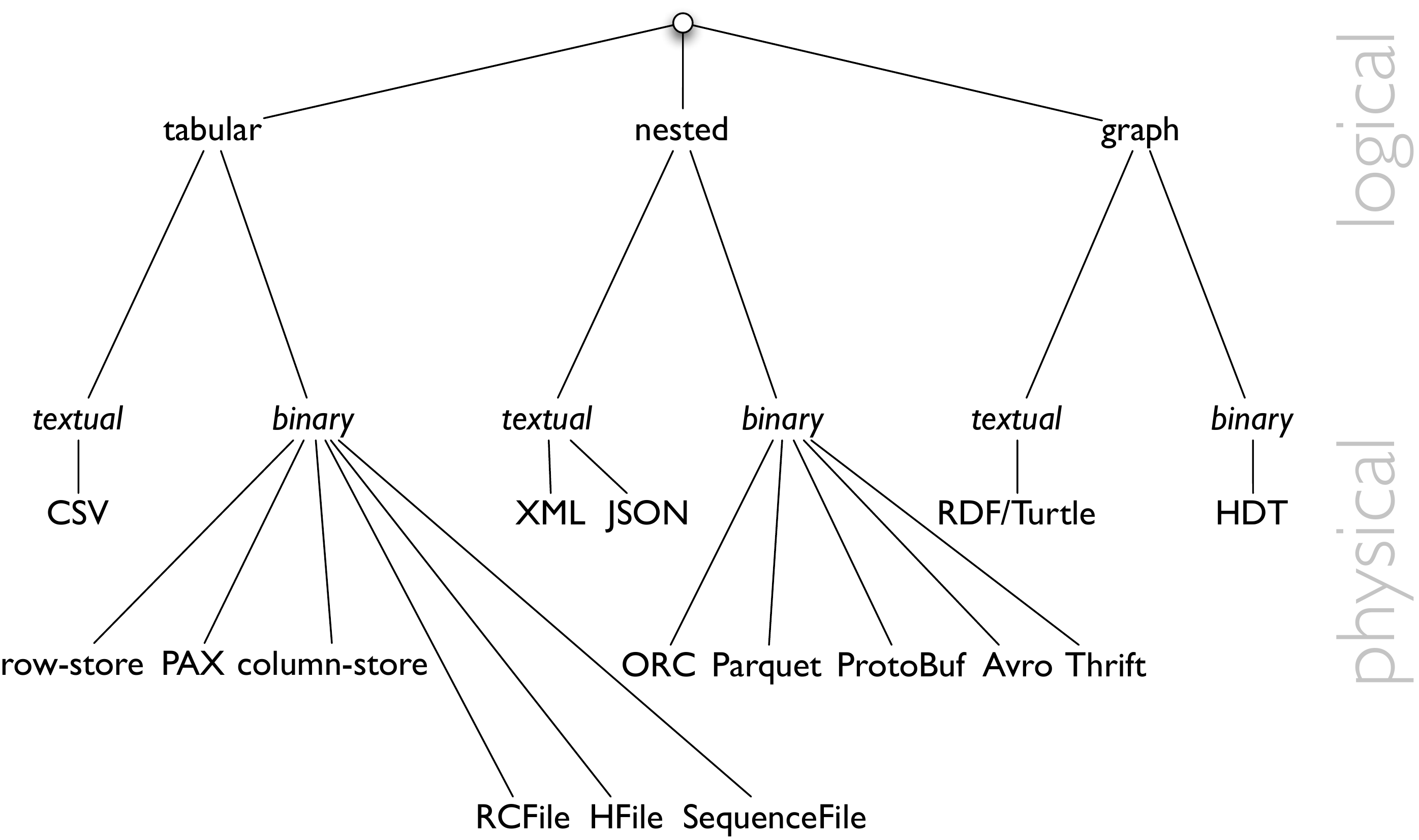}
\caption{A non-exhaustive, lightweight taxonomy for logical and physical data 
layouts and serialisation formats commonly used in the data processing 
community.}
\label{fig:taxonomy-dl}
\end{figure}

In the following, I propose a non-exhaustive, lightweight taxonomy
for logical and physical data layouts and serialisation formats as depicted in
Fig.~\ref{fig:taxonomy-dl}.

The main point of this taxonomy is to decouple the logical from the physical 
level. While for the human user the logical level is of importance, from a
software and systems perspective the physical level dominates. There are cases,
however, where the abstraction is leaking and the user is forced to accommodate.

Take, for example, best practices concerning NoSQL data modeling\footnote{As 
found in the blog post ``NoSQL Data Modeling Techniques'' via 
\url{http://highlyscalable.wordpress.com/2012/03/01/nosql-data-modeling-techn
iques/}}: with a wide-column store, such as HBase, one can easily get into a
situation where one must take into account the physical location of the data
in order to avoid performance 
penalties\footnote{\tiny \url{http://stackoverflow.com/questions/10806955/hbase-schema
-key-for-real-time-analytics-solution}}.

Also, the choice of the serialisation format (for example, textual vs. binary) 
can have severe implications, both in terms of performance and maintenance.
Look at a case where one decides to use JSON as the wire format in contrast to,
say, Avro. In the former case, one can debug any document simply by issuing a 
command on the shell like \texttt{cat datafile.json | more} while with Avro 
more 
specialised tooling is necessary. On the other hand, one can probably expect a 
better I/O performance and disk utilisation with a binary format such as Avro,
compared to JSON.

Now we're already entering the discussion of the impact of choices we make 
concerning how the data is laid out. Let's jump right into it.

\section{Impact on Data Processing at Scale}
\label{sec:ldp}
There are two schools of thought concerning the organisation of data units:
data \emph{normalisation}, and data \emph{denormalisation}. The former
wants to minimise redundancy, the latter aims to minimise assembly. Both
have their own built-in assumptions, characteristics and use cases:

\paragraph{Normalised data \ldots} 
\begin{itemize}
	\item As data items are not redundant, data consistency is 
relatively easy to achieve compared to denormalised data.
	\item When updating data in place one only has to deal with it once and
not in multiple locations.
	\item Storage is efficiently used, that is, it takes up less disk space.
\end{itemize}

\paragraph{Denormalised data \ldots} 
\begin{itemize}
	\item The access to data units is fast as no joins are necessary; the
data can be considered to be pre-joined.
	\item As it provides an entity-centric view, it is in general more 
straight-forward to employ automated sharding of the data.
	\item Due to keeping multiple copies of data items or fragments thereof
around, it requires typically a multitude more space on disk than normalised 
data.
\end{itemize}
In Table~\ref{tab:ndcomparison} I'm providing a comparison and summary of the
two different ways to handle data including typical examples of workloads and
technologies concerning use cases.
\begin{center}
\begin{table}
\caption{A comparison of normalised vs. denormalised handling of data on the 
logical and physical level across SQL and NoSQL data management systems.}
\label{tab:ndcomparison}
\begin{tabular}{l @{\hskip 2mm} p{0.30\linewidth} @{\hskip 5mm}
	            p{0.40\linewidth} @{\hskip 2mm}}
& \textbf{NORMALISED}& \textbf{DENORMALISED}\\
\hline
\hline
\emph{characteristics}&
Each data item is stored exactly in one place.&
The data items are repeated as needed.\\
\hline
\emph{advantages}&
Built-in data consistency and storage-efficiency.&
Fast, entity-centric data access without the need for joins.\\
\hline
\emph{disadvantages} &
Joints are costly and hard to implement (especially distributed).&
Inflexible and storage-hungry.\\
\hline
\emph{workloads} &
OLTP, write-intensive&
OLAP, read-intensive\\
\hline
\emph{examples} &
Classical, textbook relational database modelling&
Wide-column datastores (HBase, Cassandra), document-oriented 
datastores (MongoDB, CouchDB), key-value datastores (Redis, Memcached, etc.),
graph databases (Neo4j, RDF stores), large-scale relational databases\\
\hline
\hline
\end{tabular}
\centering
\end{table}
\end{center}
Allow me a side remark relative to the ongoing and tiring debate SQL vs. NoSQL: 
it turns out that the focus on SQL as the representative of the evil is really a
rather backward view. As stated in many places all over the
Web\footnote{For example, see the blog post 
\url{http://gigaom.com/2013/02/21/sql-is-whats-next-for-hadoop-heres-whos-doing-
it/}} many open source projects and commercial entities are introducing SQL
bindings or interfaces on top of Hadoop and NoSQL datastores. 

This is quite understandable, given the huge number of deployed (business 
intelligence) tools that natively speak SQL and of course the many people out 
there trained in this language.

\textbf{Joining the dots.} We are now in a position to wrap up on the 
impact of choices we make concerning how the data is laid out: one dimension of
freedom is the choice how we organise the data: normalised vs. denormalised. 
The second choice we have is concerning the physical data representation.
Interestingly, some systems are more rigid and upfront with what they support,
expect or allow. While, for example, in the Hadoop ecosystem it is entirely up
to you how you serialise your data---and depending on your requirements and
the workload you might end up with a different result---traditional RDBMS are
much more restrictive. Seldom you get to choose the physical data layout and
the logical layout is hard-coded anyways.

Coming back full circle to the initial Fig.~\ref{fig:data-layers} one should,
however, not underestimate the \emph{User Interface} level. At the end of the
day the usability, integrability and user familiarity of this level can be
the reason why some data management systems may have a better chance to survive
than others. Last but not least, one should take into account the emerging
\emph{Polyglot 
Persistence}\footnote{\url{http://martinfowler.com/bliki/PolyglotPersistence.htm
l}} meme that essentially states that one size does not fit it all concerning
data storage and manipulation. I suggest embracing this meme together with Pat
Helland's advice~\cite{Helland:ACMQ1}: ``In today's humongous database systems, 
clarity may be relaxed, but business needs can still be met.''

\section{Acknowledgements}
\label{sec:ack}
I'd like to thank Eric Brewer, whose RICON2012 keynote motivated me to write up
this short note. His keynote is available via \url{https://vimeo.com/52446728} 
and more than certainly worth watching it in its entirety. 

\newcommand{\etalchar}[1]{$^{#1}$}

\end{document}